\documentclass{article}
\usepackage{amsfonts}
\usepackage{amssymb}
\usepackage{pst-all}

\def\fpf{fixed-point free\ }

\title{An Algorithm to List All the Fixed-Point Free Involutions on a Finite Set}
\author{Cyril Prissette \\
\\
Laboratoire de Sondages Electromagnétiques\\
de l'Environnement Terrestre - UMR 6017\\
\\
Institut des Siences de l'Ingenieur de Toulon et du Var\\
Avenue Pompidou. B.P. 56\\
83162 La Valette Cedex - France\\
\\
Email : prissette@univ-tln.fr\\
}

\begin{document}  

\maketitle

\section*{Abstract}
A fixed-point free involution on a finite set $S$ is defined as a bijection $I : S \to S$ such as $\forall e \in S, I(I(e)) = e$ and $\forall e \in S, I(e)\ne e$.

In this article, the fixed-point free involutions are represented as partitions of the set $S$, and some properties linked to this representation are exhibited.

Then an optimal algorithm to list all the fixed-point free involutions is presented. Its soundess relies on the representation of the fixed-point free involutions as partitions.

Finally, an implementation of the algorithm is proposed, with an effective data representation. 

\section*{Keywords}
Algorithm, Fixed-point free involutions, Partitions, Recursion

\section{Introduction}
A fixed-point free involution involution on a finite set is a function which can be defined as follows :
$$\forall e \in S, I(I(e)) = e$$
$$\forall e \in S, I(e)\ne e$$

Some recent cryptanalysis methods are based on fixed-point free involutions on finite sets. Indeed, such functions can be seen as mixing functions with a structural weakness, which make them trivially invertible.

Such functions can be used instead of a cryptographically robust functions, in ordrer to study the behaviour of a cryptographyc algorithm \cite{POI06}. An other possible cryptographic attack is to find weak keys, such as the algorithm is equivalent to a fixed-point free involution \cite{PRI04}.

Obviously, an algorithm can be use to list all the permutations $\pi$ of the finite set $S$  \cite{DIJ97}. For each permutation, $I$ as $\forall e \in S, I(e) = \pi_e$, it is easy to check the constraint of involution ($\forall e \in S, I(I(e)) = e$ ) and the absence of fixed-point.

However, cryptography use large finite sets and using an algorithm to list all the permutation is a waste of time, because the number of fixed-point free involutions on a set roughly equals the square root of the number of permutations on the set.

In the first part of this article, some properties of the \fpf involutions are presented and a quick proof is given for each of them. Then, in a second part, an algorithm to list all \fpf involutions on a finite set is described. Finally, an effective data representation is proposed in an example of implementation of the algorithm.

\section{Properties of the Fixed-Point Free Involutions}
\subsection{Fixed-Point Free Involutions and Partitions}
Let $I$ be a \fpf involution on $S$ and define $P_I$ as follows :
$$P_I = \{ \{ e, I(e) \}, \forall e \in S \}$$
Obviously, as $I$ is a \fpf involution , $e$ and $I(e)=e'$ generate the same subset $\{e, I(e)\} = \{I(I(e)), e')\} = \{ I(e'), e' \} = \{ e', I(e') \}$.
Every element of $S$ is a element of a single element of $P_I$. Thus, the \fpf involution $I$ defines the set $P_I$ as a partition of $S$, such as the cardinality of every subset of $P_I$ is $2$.

Conversely, given a partition $P_I$ of $S$, such as the cardinality of every element of $P_I$ is $2$, the involution $I$ can be defined as follows :
$$\forall \{e,e'\} \in P_I, \left\{   \begin{array}{c} I(e)=e' \\ I(e')=e \end{array} \right. $$
As every element of $S$ is an element of a single element of $P_I$, then $I(I(e))=I(e')=e$.
Moreover, as the cardinality of every element of $P_I$ is $2$, thus $\forall_{e \in S}, I(e) \ne e$. So $I$ is the \fpf involution defined by $P_I$. 

The \fpf involution $I$ can be represented in a single way as the partition $P_I$. This property will be use to represent \fpf involution in a convenient way.
\subsection{Fixed-Point Free Involutions and Union}
Let $I$ be a \fpf involution on $S$. Let $i$ and $j$ be such as $i \notin S$ and $j \notin S$ and $i\ne j$. 
Considering the partition $P_I$ of $S$, associated to $I$, $P_{I'} = P_I \cup \{\{i,j\}\}$ is a partition of $S \cup \{i,j\}$ and the cardinality of every subsets is $2$.
So $P_{I'}$ can be used to represent a \fpf involution $I'$ on $S \cup \{i,j\}$, defined as follows :
$$\left\{\begin{array}{c} \forall e \in S, I'(e)=I(e) \\ I'(i)=j \\ I'(j)=i \end{array} \right. $$

The main idea of the algorithm is to build a \fpf involution, with a \fpf involution on a smaller set.
\subsection{Fixed-Point Free Involutions and Bijections}
Let $I$ be a \fpf involution on $S$,\\
Let $B$ be a bijection from $S$ to $S'$\\
Let's define $I'$ from $S'$ to $S'$ as follows :
$$I'(e') = B \circ I \circ B^{-1} (e')$$ 
Let's prove that $I'$ is a \fpf involution. First, Let's prove that $I'$ is an involution.
$$I'(I'(e'))=I' \circ B \circ I \circ B^{-1} (e')$$
$$\Leftrightarrow I'(I'(e'))=B \circ I \circ B^{-1} \circ B \circ I \circ B^{-1} (e')$$
$$\Leftrightarrow I'(I'(e'))=B \circ I \circ I \circ B^{-1} (e')$$
$$\Leftrightarrow I'(I'(e'))=B \circ B^{-1} (e')$$
$$\Leftrightarrow I'(I'(e'))=e'$$
So $I'$ is an involution.

Now, let's prove that $I'$ is fixed-point free. If $I'(e') = e'$, then 
$$B^{-1} \circ B \circ I \circ B^{-1} (e')=B^{-1} (e')$$
$$\Leftrightarrow I \circ B^{-1} (e')=B^{-1} (e')$$
$$\Leftrightarrow I (e)=e \mbox{ with } e=B^{-1}(e')$$
However, $I$ is fixed-point free. So the previous equality is false, and $I'$ is a \fpf involution.

This property is useful to build \fpf involution on any set with an even cardinality : one can build a \fpf involution on a simple set with the same cardinality, then use a bijection to map this \fpf involution onto the wanted set.

Without loss of generality, $S$ is defined as $\{1, 2, 3,..2n\}$ until the end of this article.

\section{Algorithm}

\subsection{Bijections Family}
For the purpose of the algorithm, a family of bijections from $S=\{1,2,..,2k\}$ to $\{2,..,2k+2\}\setminus i$ is needed, with $i$ in $\{2,2,..,2k+2\}$.

Although many families of bijections may be used, the following one is chosen : 
$$B_{i} (x) = \left\{   \begin{array}{c} x+2 \mbox{ if } x+1\ge i \\ x+1 \mbox{ if } x+1<i \end{array} \right.$$
Every element of the family is an easy-to-compute, easy-to-invert, increasing function. None of these properties is mandatory; however, they are useful for saving time and space.

Obviously, the inverse function of $B_{i}$ is :
$$B_{i}^{-1} (x) = \left\{   \begin{array}{c} x-1 \mbox{ if } x+1\le i \\ x-2 \mbox{ if } x+1>i \end{array} \right.$$

\subsection{Presentation of the Algorithm}
The goal of the algorithm is to construct the set of the \fpf involutions on the finite set $S=\{1,2,..,n\}$, with even cardinality. The main idea is to start with the simple \fpf involution represented by the partition $\{\{\}\}$ then, the size of the set is increased using the "Union Property" and the "Bijection Property".

This is a recursive process : given a \fpf involution on $\{1,..,k\}$, a bijection is used to get an \fpf involution on $\{2,..,2k+2\}\setminus i$, then the union property is used to add the set $\{1,i\}$ and get a \fpf involution on $(\{2,..,2k+2\}\setminus i)  \cup \{1,i\} = \{1,..,k\}$.

Here is a description of the algorithm, as a recursive function.

\begin{center}
\begin{itemize}
	\item[]{\bf function} fpfi(n,S)
	\item[] {\it // n : cardinal of the final set}
	\item[] {\it // S : current set (initial value = $\varnothing$) } 
	\item[]{\bf if} ($|S|=n$) {\bf then}
		\begin{itemize}
		\item[]output $S$
		\end{itemize}
	\item[]{\bf else}
		\begin{itemize}
		\item[]{\bf for i=$2$ to $n$}
			\begin{itemize}
			\item[]$S'=\{\{1,i\}\}$
			\item[]{\bf forall $\{e,e'\} \in S$}
				\begin{itemize}
				\item[]$S'=S' \cup \{ \{ B_i(e), B_i(e') \} \}$
				\end{itemize}
			\item[]{\bf end forall}
			\item[]fpfi(n,S')
		\end{itemize}
		\item[]{\bf end for}
		\end{itemize}
	\item[]{\bf end if}
	\item[]{\bf end}
\end{itemize}
\end{center}

\subsection{Example}
Here is a quick description of the building of one of the involutions on the set $\{1,..,6\}$, knowing an involution on the set $\{1,..,4\}$, for example the involution $I$ such as :
$$P_I = \{ \{1,3\},\{2,4\}\}$$

There are $5$ involutions built on $I$, each of them includes $\{1,i\}$ with a different value of $i$ in $\{2,..,6\}$. For each of these involutions, the associated partition is built using $I$ and $B_i$.

For example, for $i=5$, the elements of the partition are :
\begin{itemize}
\item the element $\{1,5\}$.
\item the elements built with $B_5$ and $P_I$ : 
\begin{itemize}
\item from $\{ 1, 3 \}$, compute $\{ B_5(1), B_5(3) \} = \{ 2, 4 \}$ 
\item from $\{ 2, 4 \}$, compute $\{ B_5(2), B_5(4) \} = \{ 3, 6 \}$ 
\end{itemize}
\end{itemize}

The resulting involution is associated with the set of these three sets :
$$\{ \{1,5\},\{2,4\},\{3,6\}\} $$

The following tree shows how some of the \fpf involution on $\{1,..,6\}$ are built.


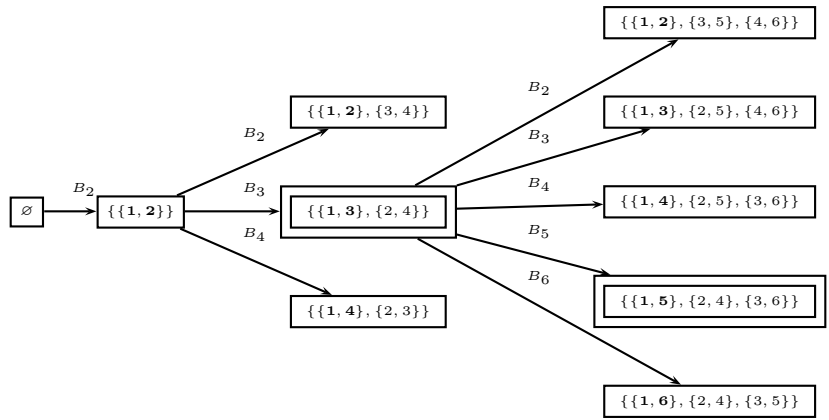
\begin{center}
{\tiny
\psset{arrows=->}
\pstree[treemode=R,levelsep=20ex]
{\TR{\psframebox{$\varnothing$}}}
{\pstree[treemode=R,levelsep=40ex]{\TR{\psframebox{$\{ {\bf \{1,2\}}\}$}} \taput{$B_2$}}
	{
	\TR{\psframebox{	$\{ {\bf \{1,2\}}, \{3,4\}  \}$		}} \taput{$B_2$} 
	\pstree[treemode=R,levelsep=60ex]{\TR{\psframebox{\psframebox{	  $\{ {\bf \{1,3\}}, \{2,4\}  \}$         	}}} \taput{$B_3$} }
		{
	        \TR{\psframebox{	$\{ {\bf \{1,2\}}, \{3,5\}, \{4,6\}  \}$	}} \taput{$B_2$}
                \TR{\psframebox{	$\{ {\bf \{1,3\}}, \{2,5\}, \{4,6\}  \}$	}} \taput{$B_3$}
		\TR{\psframebox{	$\{ {\bf \{1,4\}}, \{2,5\}, \{3,6\}  \}$	}} \taput{$B_4$}
		\TR{\psframebox{\psframebox{	$\{ {\bf \{1,5\}}, \{2,4\}, \{3,6\}  \}$	}}} \taput{$ B_5$}
		\TR{\psframebox{	$\{ {\bf \{1,6\}}, \{2,4\}, \{3,5\}  \}$	}} \taput{$B_6$}
        	}
	\TR{\psframebox{	$\{ {\bf \{1,4\}}, \{2,3\}  \}$		}} \taput{$B_4$} 
    	}
}
}
\end{center}

\section{Implementation}

\subsection{Data Representation}
As previously shown, a \fpf involution $I$ on the set $S$ can be represented as a partition $P_I$ of $S$ such as the cardinality of every element of $P_I$ is $2$.

Let's $\mu(P_I)$ be the set of elements of $S$ defined as follows :
$$\mu(P_I) = \{ min(i,j), \forall_{\{i,j\} \in P_I} \}$$

The algorithm represents the partition $P_I$ as an 2n-element array $T$. 

$$\left\{ \begin{array}{ll} 
\forall_{0\le k<n},& T[2k] \in \mu(P_I)  \\ 
\forall_{0\le k<n-1},& T[2k] > T[2k+2] \\
\forall_{0\le k<n-1},& T[2k+1] = I(T[2k])
\end{array} \right.$$

Simply speaking, the odd-indexed elements of $T$ are, decreasingly sorted, the set of the lowest elements of each set of the partition $P_I$. The even-indexed elements of $T$ are the values associated by $I$ to the odd-indexed elements.

\subsubsection{Example}
Let $I$ be the \fpf involution on $\{1..6\}$ such as $I(x)=7-x$. This \fpf involution is represented by the partition $P_I$ :
$$P_I = \{ \{2,5\}, \{3,4\}, \{6,1\} \}$$
With this partition, $\mu(P_I)$ is defined as :
$$\mu(P_I) = \{2, 3, 1\}$$
This partition $P_I$ (and so the \fpf involution $I$) is represented as the array $T$ :
$$T =\lbrack 3,4,2,5,1,6\rbrack$$

\subsection{Union operator}
The purpose of the proposed data representation is to speed up the calculation of the Union operator. Actually, with this representation, the bijection do not destroy the order of the element, and the new couple $(1,I(1))$ is simply merged at the end of the array.

In many practical implementations, it can be effectively done by allocating an array as large as the size of the set, and recursively filling the array from left to right.

An example of such an implementation is given in Appendix A.

\section{Conclusion}
The algorithm presented in this article was designed to fit cryptographic needs of an effective algorithm to list all the fixed-point free involutions on a finite set.
However, the use of such functions is not restricted to cryptographic researches, and the algorithm is generic.

\bibliography{}

%
%
%
%

\end{document}